# Predictive modeling of ion migration induced degradation in perovskite solar cells

Vikas Nandal and Pradeep R. Nair

Department of Electrical Engineering,

Indian Institute of Technology Bombay, Mumbai-400076, India

Email: nknandal@iitb.ac.in, prnair@ee.iitb.ac.in

*Abstract* — **With excellent efficiencies being reported from multiple labs across the world, device stability and the degradation mechanisms have emerged as the key aspects that could determine the future prospects of perovskite solar cells. However, the related experimental efforts remain scattered due to the lack of any unifying theoretical framework. In this context, here we provide a comprehensive analysis of ion migration effects in perovskite solar cells. Specifically, we show, for the first time, that (a) the effect of ionic charges is almost indistinguishable from that of dopant ions, (b) ion migration could lead to simultaneous improvement in Voc and degradation in Jsc - a unique observation which is beyond the realm of mere parametric variation in carrier mobility and lifetime, (c) champion devices are more resilient towards the ill effects of ion migration, and finally (d) we propose unique characterization schemes to determine both magnitude and polarity of ionic species. Our results, supported by detailed numerical simulations and direct comparison with experimental data, are of broad interest and provide a much needed predictive capability towards the research on performance degradation mechanisms in perovskite solar cells.**

## I. Introduction

Organic-inorganic hybrid Perovskites have emerged as a promising candidate for photovoltaics due to its excellent opto-electronic properties such as large diffusion length [1–3], large absorption coefficient [4], and flexibility in band gap tuning [5–8]. Methyl ammonium lead tri-iodide ($MAPBI_3$) based solar cells already report power conversion efficiencies of 21.8 % and shows excellent improvement as compared to CdTe, CIGS, and c-Silicon solar cells [9]. Low temperature solution processing, low cost raw material, and relative lack of defects are some of the other appealing attributes of this emerging class of devices.

Given the impressive gains in power conversion efficiency, it is natural that long time stability becomes the main research focus of the community. In general, degradation in solar cells is mainly manifested by a decrease in the performance metrics like $J_{sc}$, $V_{oc}$, FF, and hence the efficiency. Perovskite solar cells are no different and a degradation in the above mentioned parameters are routinely reported in literature. Further, there has been significant research to explore effects like hysteresis [10–12], giant switchable photovoltaic effect [13], giant dielectric constant [14], enhanced transistor performance at low temperature [15], photo-induced phase separation [16], self-poling [17], and electric field driven reversible conversion between $PbI_2$ and $MAPbI_3$ [18]. Many of these studies indicate that ion migration could be one of the dominant phenomena while the debate is yet to settle on the specifics of ion species that migrate and its origin. Theoretical [19–21] and experimental [18,22,23] studies indicate that $I^-$ and $MA^+$ could be the migrating species. However, the dominant mode of ion migration (i.e., through bulk crystal vs. grain boundaries) is still an unresolved topic. Moreover, recent reports indicate many surprising trends on perovskite degradation which include an improvement in $V_{oc}$ with a simultaneous degradation in $J_{sc}$ [24,25]. These trends indicate that the degradation phenomena in such solar cells are more complex and requires detailed modeling.

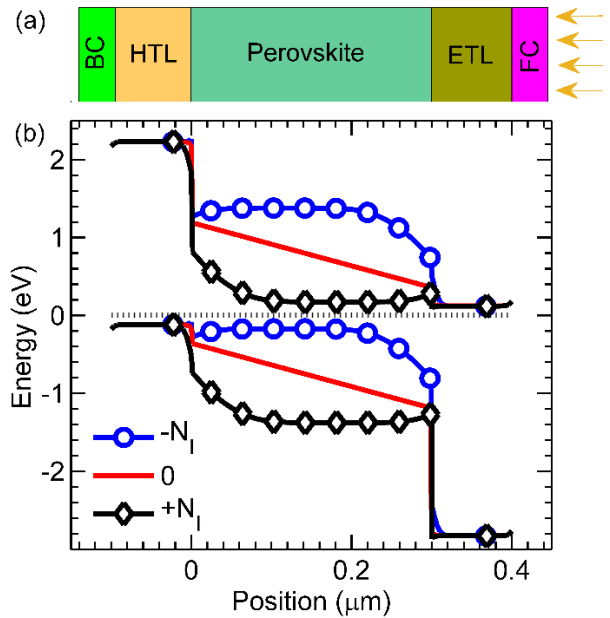

**Fig. 1.** Schematic of model system with planar PIN device architecture (a). Part (b) denotes the energy band diagram of the device with and without bulk ion charge density of $\pm N_I = 10^{17} cm^{-3}$. Depending on the polarity of the ions, the original PIN configuration changes to PPN or PNN scheme.

In this manuscript, we provide a detailed mapping of ion migration induced performance degradation of Perovskite solar cells (schematic of the cell is shown in Fig. 1a). In particular, we show that ion migration affects the device electrostatics in a non-trivial manner which could explain some of the puzzling phenomena described earlier. Our results indicate that ion migration can, surprisingly, lead to $V_{oc}$ improvement while, in general, a degradation in device performance is usually anticipated. Further, we propose novel schemes to extract the both the density of ionic charge and its polarity from device current-voltage characteristics. Indeed, our results indicate that $V_{oc}$ of champion devices are less likely to be affected by ion migration, and the effect is also critically influenced by the contact layer properties as well.

The manuscript is arranged as follows: We first provide a theoretical analysis on the influence of ions on device electrostatics, and then its effects on perovskite solar cell performance. The model predictions are then validated through detailed numerical simulations and direct comparison with available experimental data.

## II. EFFECT OF IONIC CHARGE ON SOLAR CELL PERFORMANCE

Migration of ionic species is a well-documented topic in the field of inorganic semiconductor devices. For example, the FET device reliability vastly improved due to the detailed understanding of Na$^+$ ion diffusion in gate insulators [26,27]. As such both diffusion and drift (due to an electric field) can contribute to the migration of ions. To understand the effect of ionic charges on solar cell performance, the ionic charge density needs to be accounted in the Poisson's equation as follows

$$\nabla \cdot \varepsilon \nabla \phi = -q(p - n + N_d - N_a \pm N_I), \quad (1)$$

where $\phi$ is the potential, $\epsilon$ is electrical permittivity of medium, $q$ is electronic charge, $p$ and $n$ are hole and electron densities, $N_a$, $N_d$, and $N_I$ are acceptor, donor, and bulk ionic charge densities, respectively. Here the carrier densities, n and p, depends on the electrostatic potential and the quasi fermi levels – not per se on the doping level (however, under charge neutrality conditions, the majority carrier density will be the same as effective doping levels). Hence, it is evident from the above equation that ionic charges are indistinguishable from that of ionized dopants – even though they do not provide any mobile charge carriers on their own. In such a scenario, mobile carriers, if any present, will be sourced from elsewhere, for example, other regions or the contacts. In fact, the origin or source of such mobile carriers is not predicted by Eq. (1).

Accordingly, our detailed simulations show that the effect of ionic charge is almost indistinguishable from ionized dopants even for carrier selective P-I-N device configurations. Hence, depending on the polarity of ionic species, an original P-I-N structure changes to either P-N-N or P-P-N due to ion migration (see Fig. 1b).

Based on the above arguments, we now develop an analytical model to predict the effect of ion migration on solar cell performance. For this we consider a simple perovskite solar cell (P-I-N configuration) as the model system. The $V_{oc}$ of a solar cell, by definition, is the separation between the quasi-Fermi levels at open circuit conditions. Accordingly, we have

$$V_{OC} = \frac{kT}{q} \ln\left(\frac{np}{n_i^2}\right) \quad (2)$$

where $n$ and $p$ are the electron and hole densities, respectively, and are assumed to be spatially uniform in the active layer ($n_i$ is the intrinsic carrier concentration). At $V_{oc}$, detailed balance indicates that the rate of generation of carriers ($G$) should be the same as the rate of recombination ($R$) and is given as

$$G = R = (np - n_i^2)/(\tau_n(p + p_1) + \tau_p(n + n_1)) +$$
$$B(np - n_i^2) + (A_n p + A_p n)(np - n_i^2), \quad (3)$$

where the first term in the RHS denotes the trap assisted recombination, the second term denotes the radiative recombination, and the third term denotes the Auger recombination. Here, B and $(A_n, A_p)$ are the radiative and Auger recombination coefficients, respectively. The photo-generated excess minority carrier density ($\Delta n$ or $\Delta p$) can be expressed in terms of the effective lifetime ($\tau_{eff}$) due to the above recombination mechanisms as $\Delta n \sim G\tau_{eff}$. Specifically, eq. (3) indicates that the effective lifetime for carriers in the intrinsic active region (i.e., for PIN structure with $n \sim \Delta n = \Delta p \sim p$, $\tau_n = \tau_p = \tau$, and $A_n = A_p = A$) is given as

$$\frac{1}{\tau_{eff,p}} = \frac{1}{2\tau} + Bn + An^2. \quad (4)$$

However, eq. (1) indicates that ion migration changes the effective doping density and hence changes the majority carrier density in the active layer. Hence, for negatively charged ions of density $N_I$, we have $p = N_I$ and the effective lifetime for minority carriers (i.e., electrons) is given as

$$\frac{1}{\tau_{eff,d}} = \frac{1}{\tau} + BN_I + AN_I^2. \quad (5)$$

Accordingly, the $V_{oc}$ for pristine devices (i.e., PIN structure) is given as

$$V_{OC,p} = \frac{2kT}{q} \log\left(\frac{G\tau_{eff,p}}{n_i}\right), \quad (6)$$

while that for degraded devices (with $n \approx \Delta n$, and $p = N_I$) is given as

$$V_{OC,d} = \frac{kT}{q} \log\left(\frac{GN_I \tau_{eff,d}}{n_i^2}\right). \quad (7)$$

Equations (4) - (7) predicts the $V_{OC}$ of the device before and after ion migration. As such these equations are quite general and are applicable even in the presence of significant interface recombination by modifying the bulk recombination as $1/\tau = 1/\tau_B + S/l$, where $\tau_B$ is the bulk carrier lifetime and $S$ is the carrier recombination velocity due to traps at ETL/Perovskite and HTL/Perovskite interface, and $l$ is the thickness of the perovskite layer. These equations are based on the assumption that the dominant effect of ion migration is in the electrostatics of the device. However, ion migration could create more traps in the active layer. Although this effect is not explicitly discussed in this manuscript, eqs. (4) - (7) predict the $V_{oc}$ in such a scenario with appropriate $\tau$. Further, eq. (7) is valid for both polarities of ions, irrespective of the specific assumptions made in the derivation.

## III. MODEL PREDICTIONS

The analytical model described in the previous section makes some interesting predictions which are listed below-

a) **$V_{oc}$ improvement**: The theoretical limit for open circuit voltage for perovskite based solar cells is 1.17V (for an assumed band gap of 1.55eV). This limit is due to radiative and Auger recombination processes [28]. However, the reported $V_{OC}$ values in literature are significantly lower than the theoretical limits which indicates that trap assisted recombination (could be bulk or interface) is the dominant mechanism [24,29,30]. Under such conditions, we have $\tau_{eff,p} = 2\tau$ and $\tau_{eff,d} = \tau$. Accordingly, comparison of eq. (6) with eq. (7) indicates that the $V_{oc}$ of degraded devices can be better than that of the pristine devices if $N_I > 4G\tau$. Further, the $V_{oc}$ is expected to be increase with $N_I$ as the effective lifetime is still governed by trap assisted SRH (and hence is insensitive to $N_I$, see eq. (5)). This $V_{oc}$ improvement will reduce for larger values of $N_I$ due to increased radiative and Auger recombination (as per eq. (5)). It is interesting to note that in the absence of any increase of ionic charges during degradation, the $V_{oc}$ can improve only if the carrier lifetime increases – which might not be the scenario in most cases.

b) **$J_{sc}$ degradation**: The presence of ionic charges can affect the electrostatics of the device quite significantly as shown in Fig. 1b. Accordingly, the dominant carrier collection phenomena can change from drift to diffusion based transport. However, this need not result in any appreciable change in $J_{sc}$ as carrier collection length $l_d$ ($l_d = \sqrt{D\tau}$, $D$ is diffusion coefficient) could still be much larger than the device thickness. For very large $N_I$, the effective lifetime decreases due to increased radiative and Auger recombination (see eq. 5) and this could lead to a reduction in collection efficiency and hence the $J_{sc}$. In addition, the $J_{sc}$ can further decrease due to the unfavorable band bending at perovskite/transport layer interface. Note that in the absence of any ion migration effects, the $J_{sc}$ can decrease if parameters like carrier mobility or carrier lifetime decreases. However, a decrease in $\tau$ will result in a simultaneous decrease in $V_{oc}$, while a change in mobility will not affect the $V_{oc}$ at all. Hence a simultaneous observation of $J_{sc}$ decrease and $V_{oc}$ improvement could be a signature of ion migration effects.

c) **FF variation**: The trends for FF which is a measure of charge collection efficiency at near $V_{oc}$ conditions (to be precise, at maximum power point condition) is rather difficult to predict as compared to that of $J_{sc}$ and $V_{oc}$. In general, the FF is shown to follow the $V_{oc}$ trends for inorganic solar cells [31] and we expect similar results here as well.

d) **Efficiency**: The above discussion indicates that if ion migration is the dominant phenomena, then there could be an improvement in $V_{oc}$, and a decrease in $J_{sc}$. Accordingly, the efficiency could show interesting trends. For example, if $J_{sc}$ decrease dominates over $V_{oc}$ improvement, then the efficiency of degraded device will decrease.

## IV. MODEL VALIDATION

To test the model predictions, we performed detailed numerical simulations. In addition, we also compare our results with recent experimental results. For the numerical simulations we considered a perovskite based P-I-N solar cell. The device schematic and the energy level of various materials is provided in Fig. 1a and Fig. SS1 (supplementary materials), respectively. Uniform photo-generation is assumed in the active layer (with thickness $l = 300nm$, and $G = 4.75 \times 10^{21} cm^{-3} s^{-1}$, which corresponds to $J_{sc} = 22.8 \, mA/cm^2$). The recombination mechanisms in the perovskite layer is already mentioned in eq. 3. ETL and HTL are n and p type doped, respectively, which facilitates in the extraction of corresponding charge carriers. The Poisson and continuity equations [32] for electrons and holes are self-consistently solved to explore the effect of ionic charge on device performance. The effects of ions are explicitly considered in the electrostatics as mentioned in eq. (1). The simulation methodology is well calibrated with previous experiments [28,33] and the parameters used for this study are provided in Table S1 of supplementary materials.

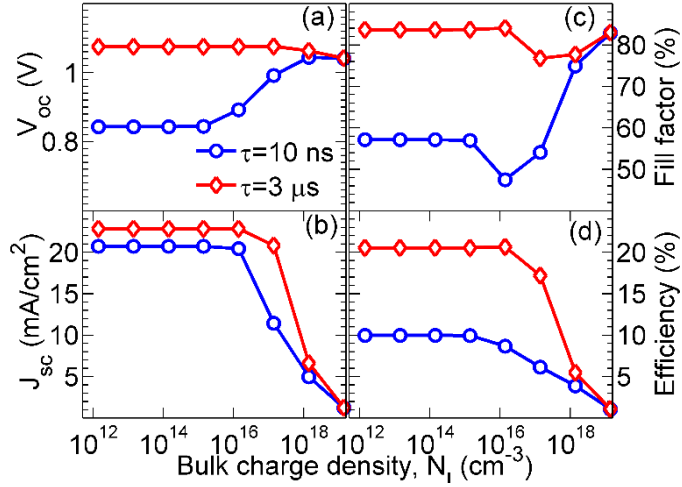

**Fig. 2**. Effect of bulk ionic charge density $N_I$ on performance matrices - (a) $V_{oc}$, (b) $J_{sc}$ (c) fill factor (d) efficiency of Perovskite solar cells. The red curve denote a device with large $\tau$, and hence a large diffusion length, while the blue curve denotes a device with small diffusion length. The results indicate that simultaneous improvement in $V_{oc}$ and degradation in $J_{sc}$ is a unique feature of ion migration effects.

Figure 2 shows the variation of $V_{oc}, J_{sc}$, FF, and efficiency as a function of ionic charge density. To explore the effects of ionic charge in detail, here we consider two cases: (a) device with diffusion length more than the perovskite thickness (assumed $\tau = 3\mu s$, which corresponds to $l_d = 1.24\ \mu m$) and (b) device with diffusion length less than the active layer thickness (assumed $\tau = 10 ns$, which corresponds to $l_d = 72\ nm$). For both cases, in addition to the above mentioned SRH recombination process, we explicitly consider radiative ($B = 3 \times 10^{-11} cm^3/s$), and Auger ($A_n = A_p = 10^{-29} cm^6/s$) [28] processes as well. The results shown here are for positive polarity of the ions, however, the trends are the same for negative ions as well (as shown by our numerical simulations, however, there could be significant differences if the ETL/HTL has dissimilar properties like doping density, dielectric constant, etc.)

Part (a) of Figure 2 shows that $V_{oc}$ of degraded device indeed improves as compared to that of the pristine device for low $\tau$, while there is no $V_{oc}$ improvement for devices with large $\tau$. These surprising results are very much in accordance with the model predictions. The minimum ionic charge density at which $V_{oc}$ shows an improvement is given by $N_I = 4G\tau$. For $\tau = 10 ns$, this correspond to $N_I \approx 10^{15} cm^{-3}$ and the $V_{oc}$ indeed shows an improvement in this case. For $\tau = 3\mu s$, the ionic charge density required for any improvement in $V_{oc}$ is of the order of $10^{17} cm^{-3}$. However, for such large ionic densities, the effective lifetime decreases due to increased radiative and Auger recombination and hence the $V_{oc}$ decreases. Specifically, eq. 5 indicates that for $N_I = 10^{17} cm^{-3}$ the radiative lifetime is given as $(BN_I)^{-1} = 0.3\mu s$, which is much lower than the assumed SRH lifetime of $\tau = 3\mu s$ and hence lowers the $V_{oc}$. The results in Fig. SS2 of supplementary materials indicate that the $V_{oc}$ improvement is expected to be significant for devices with low $\tau$. Champion devices (with excellent $V_{oc}$ or large $\tau$), quite surprisingly, are least affected by the electrostatics of ion migration. Hence, it is imperative that the active layer material quality should be excellent for ion migration tolerant performance.

The effect of ionic charge on $J_{sc}$ is shown in Part (b) of Fig. 2. For devices with PIN configuration, the charge collection efficiency $\eta$ is given as $\eta = l_c(1 - e^{-l/l_c})/l$, where $l_c$ is the drift collection length ($l_c = \mu\tau_{eff}E$, where $E$ is the built-in electric field) and $l$ is the active layer thickness [34]. The above equation predicts that the collection efficiency is 100% for $\tau = 3\mu s$, while it reduces to 88% for $\tau = 10 ns$. Note that the simulation results in Figure 2b (i.e., the ratio of $J_{sc}$ at low $N_I$) show the same trends as the above theoretical predictions. Figure 2b also shows that the $J_{sc}$ is not affected for low values of $N_I$ as the effective lifetime remains a constant (see eq. 5). However, as $N_I$ increases, the electrostatics of the device is significantly affected and the PIN scheme changes to PPN, thus changing the dominant charge collection mechanism from drift to diffusion. For very large $N_I$, the effective lifetime decreases due to the increase in radiative and Auger recombination process and hence the $J_{sc}$ degrades significantly. Note that these results are in accordance with the model predictions in Section III (b).

The fill factor variation due to ion migration is provided in Part (c) of Figure 2. Here, we find that the FF, more or less, follows the trends in $V_{oc}$. Specifically, we find that the FF is not much affected for devices with large $\tau$, while for small $\tau$ the FF shows an increase for large ionic densities. As $N_I$ increases the fraction of depletion width shared by active layer ($W_{ac}$) decreases. This improves the fill factor as the voltage dependence on charge collection efficiency decreases. Putting all together, we find that the efficiency of devices degrades due to ion migration (Fig. 2d). It is evident that the ion density required for a performance degradation in champion devices is more than that of low performance devices. This implies that the device lifetime or the stability is also expected to be better for champion devices (provided ion migration is the dominant degradation mechanism).

We now compare our simulation results with recent experiments. Recently, $V_{oc}$ improvement and $J_{sc}$ reduction during degradation was reported by Guerrero et al. [24] (see inset of fig. 3). Here the $J_{sc}$ reduces by 40% while the $V_{oc}$ improves by 150 mV during device degradation. These experimental trends are consistent with our model predictions and detailed numerical simulations. Interestingly, our numerical simulations indicate that the observed effects could be due to the presence of ionic density of $N_I \approx 10^{17} cm^{-3}$ inside the perovskite layer (under the assumption that the transport layers are similarly doped. See Fig. SS3 of supplementary materials for discussion on the effects of transport layer doping). In addition, our results also indicate that observed experimental trends could not be explained through simple parametric variations in $\mu$ and $\tau$ alone (shown in Fig. SS4, supplementary materials). For example, it would require a significant improvement in $\tau$ (a factor of around 20 to account for 150 mV improvement in $V_{oc}$ - as per eq. (6)) and dramatic reduction in $\mu$ (a factor of $10^{-2}$ to account for the 40% decrease in $J_{sc}$) to explain the observed experimental characteristics in the absence of any ion migration effects. While not impossible, such a simultaneous change in mobility and lifetime leads to a significant reduction in fill factor – which is contradictory to experimental observations (see Fig. SS5, supplementary materials). These results indicate that ion migration could be the dominant phenomena that contribute to the device degradation.

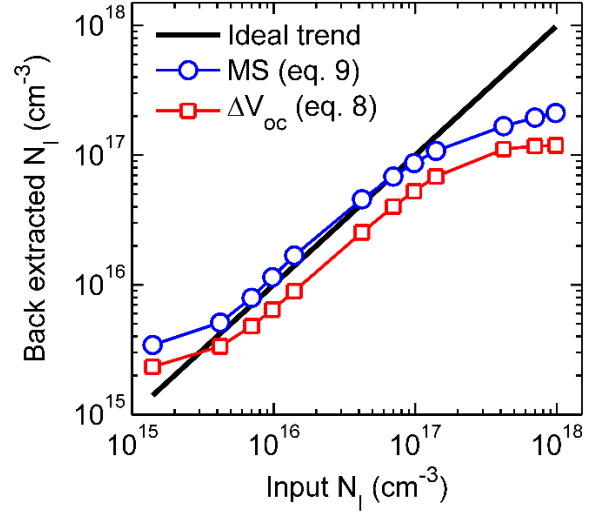

**Figure 4**. Comparison of back extracted ionic charge density from CV measurements under dark conditions (using eq. 9- circle symbols) and $V_{oc}$ improvement (using eq. 8 – square symbols). Here the x- axis denote the assumed $N_I$ for dark CV or illuminated IV numerical simulations while the y-axis denote the back extracted $N_I$.

**Back extraction of Ionic charge density**: Ionic charge density can be extracted from the $V_{oc}$ improvement ($\Delta V_{oc} = V_{oc,d} - V_{oc,p}$) observed in the degraded device. With trap assisted SRH as the dominant recombination mechanism, equations 6 and 7 indicates that ionic charge density can be back extracted as

$$N_I = 2n_i \exp\left(\frac{q\Delta V_{oc}}{kT} + \frac{qV_{oc,p}}{2kT}\right). \quad (8)$$

Figure 4 shows a comparison of the back extracted ionic density using the above equation. Here, the x-axis denote the ion density assumed in simulations while the y-axis denotes the back extracted ion density using eq. (8) (i.e., obtained from the simulated current voltage characteristics under illumination). The results indicate that eq. 8 predicts the ionic density very well. However, back extraction from $V_{oc}$ improvement is not very accurate at higher input $N_I$ because $\Delta V_{oc}$ is limited by auger recombination. To further validate our model, we employed the well-known dark capacitance voltage (CV) characteristics technique to extract ion charge density. Doping density for $P^+N$ (or $PN^+$) diode at the edge of lightly doped depletion region can be determined by using Mott-Schottky (MS) analysis and hence the back extracted ion density is given by

$$N_I = -\frac{2}{\epsilon q \left(\frac{d\left(\frac{1}{C^2}\right)}{dV}\right)} \quad (9)$$

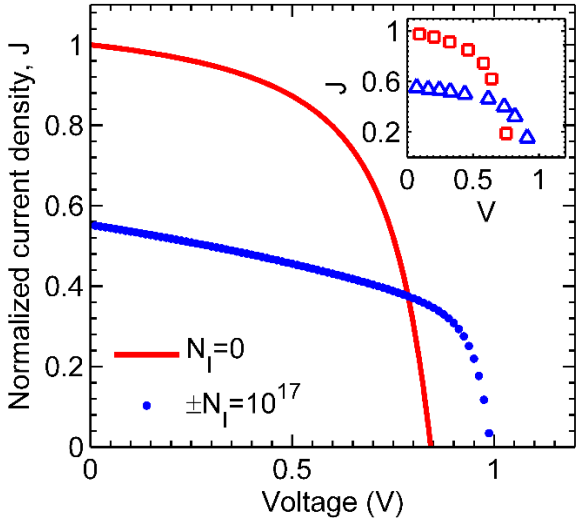

**Figure 3**. Comparison of numerical simulation of ion migration effects with experimental data from ref. [24] (see inset, open square symbols represent pristine device while the open triangles represent the degraded device). The experimentally observed $V_{oc}$ improvement and $J_{sc}$ degradation is well anticipated by numerical simulations (with $\pm N_I = 10^{17} cm^{-3}$).

## V. DISCUSSIONS

Beyond the performance degradation of perovskite solar cells, there is a significant interest among the photovoltaic community to identify the ionic species that contributes to ion migration effects. As mentioned before, both $I^-$ and $MA^+$ ions are among the likely species [18-23]. In this regard, here we discuss the possibility of finding the polarity and the magnitude (or density) of migrated ionic species in the active layer.

Here, $C$ is capacitance and $V$ is applied DC bias voltage. We find that back extracted $N_I$ from MS analysis follows similar trend and is in close agreement with input $N_I$. Therefore, $V_{oc}$ improvement from pristine to degraded device and dark CV characteristics of degraded device can be used to quantify the ion species present inside perovskite layer. In addition, eq. (8) provides a simple alternative to CV analysis for back extraction of ionic charge densities.

**Polarity of ion species**: The previous analysis indicates that CV and $V_{oc}$ improvement can be used to back extract the density of ionic species in the active layer. However, both these methods are not suited to identify the polarity of ions and this requires a novel characterization scheme. We note that the polarity of ion species dictates the electrostatics of the active layer (as evident from fig. 1b). Accordingly, depending on the polarity of the ionic species, we find that the band bending in the active layer could be near the ETL or HTL. This distinct nature of band bending helps us to identify the polarity of ions. Depending on band bending, collection efficiency of localized photo-generated electron-hole pairs near perovskite/ETL interface could be different. Therefore, we propose that quantum efficiency measurements using localized carrier generation (using photons of appropriate wavelength) can characterize the polarity of ions.

Fig. 5 compares the collection efficiencies (or normalized current density at short circuit conditions) for various scenarios with light being incident from the ETL side as $\tau$ (i.e., only the SRH lifetime) is varied while the radiative and Auger recombination rates are the same as that in Figure 2. The localized generation of photo-carriers is accounted through a spatially uniform carrier generation over a region of thickness 60nm near the ETL/perovskite interface. Here we consider three scenarios: Case A: Intrinsic perovskite, Case B: negative ions in perovskite, and Case C: positive ions in perovskite. Case A has uniform electric field in perovskite and the charge collection mechanism is due to drift. Case B has significant band bending near the ETL/perovskite interface. For the localized generation near this interface, the collection efficiency is higher as the electric field is much larger than in case A. However, for Case C the band bending is near the HTL/perovskite interface and hence the dominant collection mechanism for photocarriers generated near the ETL/perovskite interface is diffusion. For all the above scenarios, the collection length ($l_c$ or $l_d$) in Case B is larger than the collection length in Case A which is in turn larger than the collection length in Case C. As a result, the collection efficiency for case B saturates to 100% at much lower $\tau$ than for Case A because of higher electric field ETL/perovskite interface. Even for devices with large $\tau$, the collection efficiency for Case C is not 100% as the radiative recombination becomes the dominant phenomena (see Fig. SS6 of supplementary material for details). This contrasting behavior in collection efficiency is due to different band bending near ETL/perovskite interface and thus enable us to characterize polarity of ions. Therefore, EQE measurements of pristine and degraded devices can be used as a tool to identify the polarity of ions present inside active layer. Note that being a technique that relies only on device terminal characteristics, the proposed method is very well suited for fully packaged solar cells as well.

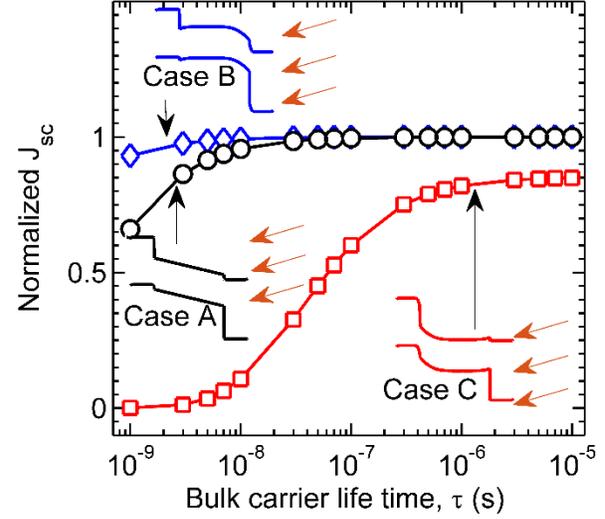

**Figure 5**. Proposed characterization scheme to identify the polarity of ions in the active layer. Insets show the EB diagrams of different cases considered: Case A ($N_I = 0$), Case B ($N_I = -10^{17} cm^{-3}$), and case C ($N_I = +10^{17} cm^{-3}$). Localized photogeneration is assumed near ETL/perovskite interface and the $J_{sc}$ is normalized to that of 100% collection efficiency. The results show that the $J_{sc}$ for the case C is significantly lower than that of case A, while the $J_{sc}$ for case B is either better than or same as that of case A. This contrasting nature of $J_{sc}$ is due to the ion polarity dependent band bending in the active layer and helps to identify the polarity of ions in the perovskite layer.

We have, so far, provided a comprehensive analysis on the effects of ion migration on performance metrics of perovskite solar cells. In addition, we have also proposed novel characterization schemes to identify the polarity and magnitude of the dominant ionic species. We now critically analyze some of the underlying assumptions to check the broad applicability of the results – (a) we have assumed that ionic species is not contributing to any further trap states in the bulk of the material. However, the model developed (eq. (7)) is equally applicable in this case as well with modified $\tau$. (b) the assumptions on ETL/HTL properties – Here we assumed that the ETL/HTL are heavily doped. Heavy doping of ETL/HTL ensures that any ion migration effects in the corresponding layers are negligible. However, there are some subtle features where ETL/HTL properties could be significant. Detailed discussions on such aspects are provided in supplementary materials (see fig. SS3). (c) assumption of uniform ionic density in perovskite layer – This is clearly not a realistic case, at least, during the initial phases of ion migration. However, this assumption aids in obtaining crucial insights and for developing the analytical model. Our numerical simulation scheme is well suited to handle non-uniform ion profiles as well and the corresponding results will be communicated later. (d) effect of ion mobility and hence any hysteresis effects – Here, we assumed that the ion profiles are static during the measurement window and there are no hysteresis effects (trapping/de-trapping effects at transport layer/perovskite interfaces). This assumption is also supported recent experimental effects [24]. Although such hysteresis effects can be accounted in our numerical simulations, the same is beyond the scope of current manuscript which provides a benchmark theory for the basic ion migration effects in perovskite solar cells.

## IV. CONCLUSIONS

To summarize, here we provided a comprehensive theoretical analysis – supported by detailed numerical simulations and experimental observations - on the electrostatic effects of ion migration in perovskite solar cells. Our predictive modeling provides physical insights into unique features observed in performance ($V_{oc}$ improvement and $J_{sc}$ degradation in comparison to pristine device) of degraded device. We find that ion migration might be the most probable physical mechanism that leads to such experimentally observed puzzling results. We indicate that champion devices are more tolerant towards ion migration effects. In addition, we have also proposed novel characterization schemes to probe the polarity and density of the migrating ionic species. Indeed, our results and the solution methodology are of broad interest and helps the community to understand the ion migration effects in isolation and thus explore the relative contribution of other mechanisms, if any, in greater detail.

**Acknowledgements:** This article is based upon work supported under the US-India Partnership to Advance Clean Energy-

Research (PACE-R) for the Solar Energy Research Institute for India and the United States (SERIIUS), funded jointly by the U.S. Department of Energy (Office of Science, Office of Basic Energy Sciences, and Energy Efficiency and Renewable Energy, Solar Energy Technology Program, under Subcontract DE-AC36-08GO28308 to the National Renewable Energy Laboratory, Golden, Colorado) and the Government of India, through the Department of Science and Technology under Subcontract IUSSTF/JCERDC-SERIIUS/2012 dated 22$^{nd}$ November 2012. The authors also acknowledges Center of Excellence in Nanoelectronics (CEN) and National Center for Photovoltaic Research and Education (NCPRE), IIT Bombay for computational facilities.

**Author Information:** All correspondence should be addressed to V.N. at nknandal@iitb.ac.in or P.R.N. at prnair@ee.iitb.ac.in. Authors declare no competing financial interest.